\documentclass[aps,prl,twocolumn,superscriptaddress,floatfix,letter,longbibliography]{revtex4-2}

\usepackage{epsfig,amsmath,amssymb,color,amsfonts,physics,microtype}
\usepackage[bookmarks=true,colorlinks,citecolor=red]{hyperref}
\usepackage{dsfont}
\usepackage{wrapfig}
\usepackage{tikz}
\usepackage{physics}
\usepackage{mathrsfs}
\usepackage[export]{adjustbox}
\usetikzlibrary{quantikz}

\usepackage{soul}

\newcommand{\cnot}{\mathfrak{\hat C}}

\begin{document}

\title{Hybrid Stabilizer Matrix Product Operator}

\author{Antonio Francesco Mello}
\affiliation{International School for Advanced Studies (SISSA), via Bonomea 265, 34136 Trieste, Italy}
\author{Alessandro Santini}
\affiliation{International School for Advanced Studies (SISSA), via Bonomea 265, 34136 Trieste, Italy}
\author{Mario Collura}
\affiliation{International School for Advanced Studies (SISSA), via Bonomea 265, 34136 Trieste, Italy}
\affiliation{INFN Sezione di Trieste, 34136 Trieste, Italy}

\begin{abstract}
    We introduce a novel hybrid approach combining tensor network methods with the stabilizer formalism to address the challenges of simulating many-body quantum systems. By integrating these techniques, we enhance our ability to accurately model unitary dynamics while mitigating the exponential growth of entanglement encountered in classical simulations. We demonstrate the effectiveness of our method through applications to random Clifford T-doped circuits and Random Clifford Floquet Dynamics. This approach offers promising prospects for advancing our understanding of complex quantum phenomena and accelerating progress in quantum simulation.
\end{abstract}

\maketitle

\paragraph{Introduction. ---}\label{s:introduction}
One of the most challenging as well crucial task in quantum physics is the simulation of the unitary dynamics of a many-body system. As a matter of fact, 
this is of critical significance for two primary reasons: first, it propels scientific advancements in areas such as condensed matter physics and quantum chemistry where large, error-corrected devices are still unavailable \cite{bravyi2024high};
second, it serves as a means to scrutinize the claims of quantum advantage put forth by state-of-the-art devices \cite{cheng2023noisy,bluvstein2024logical}. 

Indeed, as the simulation encompasses numerous qubits alongside a vast array of unitary gate operations, the level of entanglement required to encode the wave function imposes a significant constraint on contemporary state-of-the-art classical numerical techniques, such as those relying on Tensor Networks (TN) \cite{Vidal_2004,Schollwock_2011,Paeckel_2019}.
Even though TN-based techinques are able to keep only the ``relevant''  entanglement degree across the system, they are still limited by the auxiliary space dimensions $\chi$ of the ansatz. For instance, in one dimension, using the Matrix Product State (MPS) framework \cite{Schollwock_2011}, we maintain computational complexity at $O(N\chi^3)$, but, as a result, we are limited in the amount of entanglement we can possibly encode into the wave function.

Regrettably, it is a well-established fact that generic unitary evolution introduces correlations throughout the system, propagating linearly over time \cite{Calabrese_2005,RevModPhys.80.517}. This inevitably leads to a linearly increasing entanglement, resulting in an exponential rise in the MPS bond dimension, ultimately causing the breakdown of the classical simulation \cite{Läuchli_2008}.

However, in physics, our focus often lies not really on the complete evolution of the many-body wave function, but rather on the expectation values of local observables.
This remains one of the most challenging, yet fundamental,  problem in many-body physics -- basically the evaluation of $\bra{\psi} \hat U^{\dag} \hat O \hat U \ket{\psi}$ -- for a generic unitary evolution $\hat U$ and local observable $\hat O$, starting from a relatively short-range correlated state $\ket{\psi}$ (e.g. an MPS) \cite{PhysRevLett.102.240603,muller2012tensor}.

In particular, when the unitary evolution is governed by a Clifford operator ({\it cfr. Clifford \& Magic}), it is well know that the computational complexity of this task reduces from being exponential to polynomial in the number of qubits $N$ \cite{Gottesman_1997, Gottesman_1998_1, Gottesman_1998_2, Dehaene_2003, Howard_2017}.
Indeed, the actual quantum complexity of such a task relies on the interplay between the entanglement and another fundamental quantum resource dubbed non-stabilizerness (or Magic) \cite{Winter_2022,Leone_2022,niroula2024phase,lami2023quantum,mello2024retrieving,PRXQuantum.4.040317, PhysRevLett.131.180401,Heinrich_2019,Turkeshi_2023_1,fux2023entanglementmagic,tarabunga2024nonstabilizerness,PhysRevA.109.L040401,Haug_2023_2,frau2024nonstabilizerness}. 

\begin{figure}[t!]
\includegraphics[width=\linewidth]{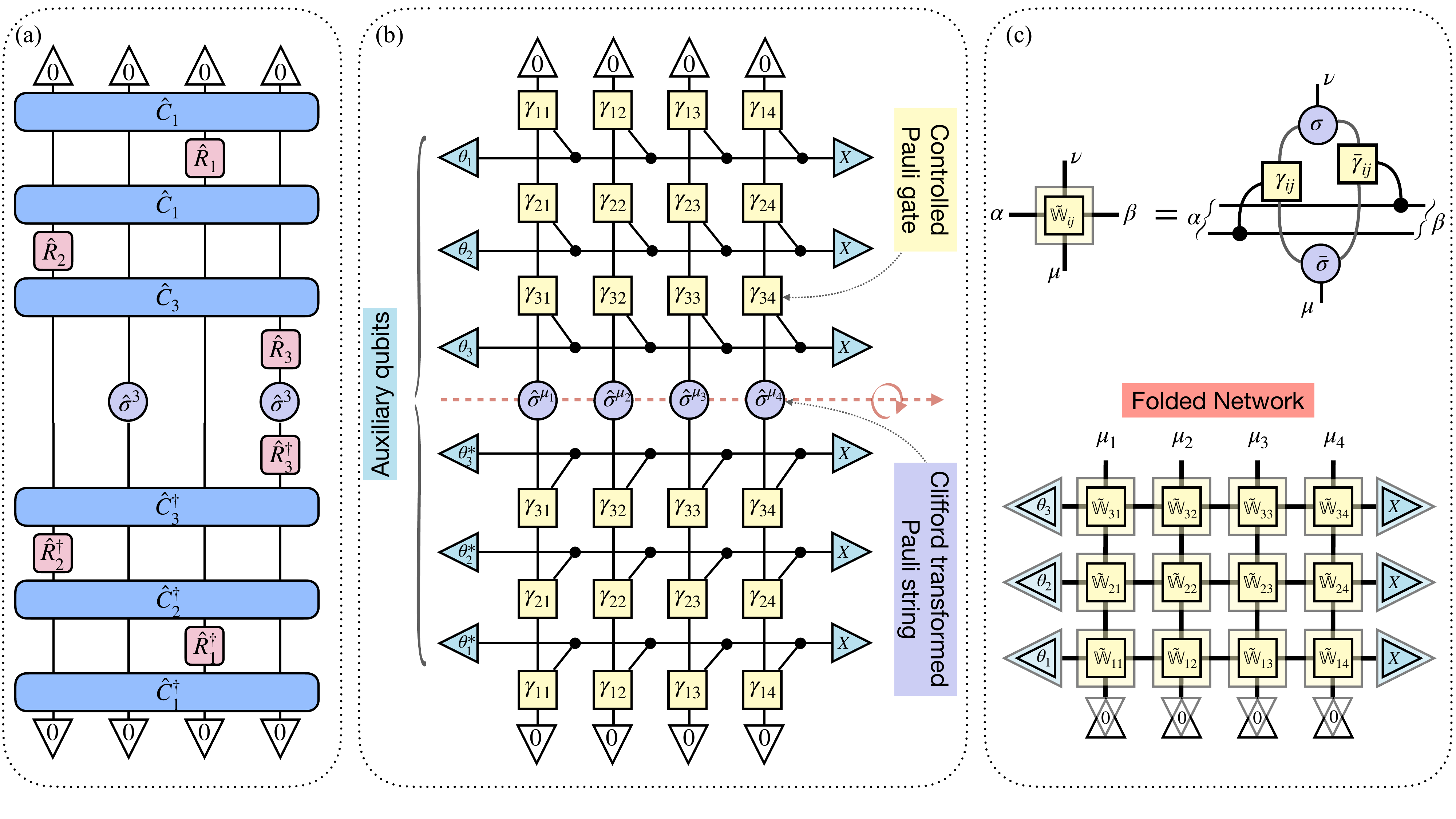}
\caption{Tensor Network contraction in the stabilizer MPO formalism. (\textbf{a}) Example of $3$ layers of Clifford and Magic operations acting on $4$ qubits initialised in $\ket{0000}$; the network is contracted with its complex conjugate to compute the expectation value of a two-point function. (\textbf{b}) The transformed network, in the stabilizer MPO formalism, exhibits a bi-dimensional grid comprising Pauli gates $\hat\sigma^{\gamma_{ij}}$ (as depicted with subscripts only in the figure), controlled by auxiliary qubits $\ket{\theta_i} \equiv \cos(\theta_i/2)\ket{0} \mp i \sin(\theta_i/2) \ket{1}$
(and $\ket{\theta^{*}_i}$), which serve to induce slight entanglement between each individual row and all columns, i.e. the physical qubits. The majority of the entanglement is in fact effectively accounted for through a straightforward transformation of the original Pauli string. Finally each row of the entire network is contracted to the not normalised state
$\ket{X} \equiv \ket{0}+\ket{1}$.
(\textbf{c}) Folded TN in the Pauli basis as a bi-dimensional grid of $\mathbb{\tilde W}_{ij}$ $4$-order tensors, whose entries coincides with those of the $\mathbb{W}$ (see main text) after discarding the coefficient $\phi_0$ and $\phi_1$ which here are absorbed in the definition of the $\ket{\theta_i}$ boundary vectors.
}
\label{fig:stabMPO}
\end{figure}

Developing smart classical algorithm able to deal with both entanglement and non-stabilizerness is of paramount importance \cite{lami2024quantum,masotllima2024stabilizer}.  
Here we develop a new strategy which use the stabilizer formalism \cite{Aaronson_2004, https://doi.org/10.48550/arxiv.1711.07848,https://doi.org/10.48550/arxiv.0807.2876} into the TN framework,
thus exploiting the best of both approaches to improve 
our ability to simulate the dynamics of many-body quantum systems.

We develop a hybrid stabilizer and TN scheme to disentangle time-evolved MPS. This approach allows us to compute exact averages of expectation values for larger times compared to standard methods, at fixed amount of resources, $\chi$. 

Specifically, we decompose the evolution into a TN evolution and a Clifford operator. We leverage the stabilizer formalism to apply the Clifford operator to the local observable,  evolve the state with TN techniques and finally compute the expectation value of the modified observable.

The paper is organized as follows: first, we introduce the basic TN and Clifford tools then, we explain our technique and finally apply it to two different setups:
(i) random Clifford $T$-doped circuits; (ii) Floquet dynamics induced by cyclically applying a series of local non-perfect kicks, followed by random $U(1)$-symmetric $N$-qubit Clifford unitaries.\\

\paragraph{Setup \& Notation. ---} \label{s:setup}
Let us consider a system with $N$ qubits, and identify the local computational basis $\{\ket{0}, \ket{1}\}$ with the eigenstates of the $\hat \sigma^3$ Pauli matrix, such that
$\hat \sigma^3 \ket{s} = (- 1)^s \ket{s}$.

Any operator acting on a single qubit can be decomposed in terms of the Pauli matrices and the identity matrix, 
i.e. $\mathcal{P}=\{\hat \sigma^{\mu}\}_{\mu=0}^{3}$, such that 
$\Tr(\hat \sigma^{\mu}\hat \sigma^{\nu}) = 2\delta_{\mu\nu}$.
Using this local basis, we can construct a generic Pauli string
as $\hat \Sigma^{\boldsymbol{\mu}}=\hat\sigma_{1}^{\mu_1} \hat\sigma_{2}^{\mu_2}
\cdots \hat\sigma_{N}^{\mu_N} \in \mathcal{P}_{N}= \mathcal{P}^{\otimes N}$,
where the boldface apex $\boldsymbol{\mu}$ identifies the set $\{\mu_1,\dots,\mu_N\}$. In subsequent discussions, whenever the subscript denoting the qubit is unnecessary, it will be omitted for the sake of notation simplicity.

The Pauli strings are therefore a complete basis for any operator $\hat O$ acting on the many-body Hilbert space $\mathcal{H} = \{\ket{0}, \ket{1}\}^{\otimes N}$, namely
$\hat O = \sum_{\boldsymbol{\mu}} O_{\boldsymbol{\mu}} \hat \Sigma^{\boldsymbol{\mu}}$ where the coefficients are given by 
$O_{\boldsymbol{\mu}} = {\rm Tr} ( \hat O \hat \Sigma^{\boldsymbol{\mu}}) /2^N$,
and we used the orthogonality condition 
${\rm Tr}(\hat\Sigma^{\boldsymbol{\mu}}\hat\Sigma^{\boldsymbol{\nu}}) 
= 2^N\delta_{\boldsymbol{\mu}\boldsymbol{\nu}}$.

In many situation the coefficient 
$O_{\boldsymbol{\mu}}\in\mathbb{C}^{4^N}$, which is an order-$N$ tensor
spanning an exponentially large space, admits an exact Matrix Product Operator (MPO) representation 
$O_{\mu_1,\mu_2,\dots,\mu_N} 
= \mathbb{O}_{1}^{\mu_1}\mathbb{O}_{2}^{\mu_2}\cdots \mathbb{O}_{N}^{\mu_N}$,
where $\mathbb{O}_{j}^{\mu_j}$, for all $\mu_j$, is a local generic matrix acting on an auxiliary space.
The dimensions of this space may change depending on the local bond, and we will often refer to it as operator bond-dimension or simply bond-dimension. 
When associated with the local operator $\hat\sigma^{\mu_j}$, 
we may construct the  operator-valued matrix 
$\mathbb{\hat O}_{j} = \sum_{\mu}\mathbb{O}_{j}^{\mu}\hat\sigma^{\mu_j}$,
such that we get the compact notation for the original operator itself
$\hat O = \mathbb{\hat O}_{1}\mathbb{\hat O}_{2}\cdots \mathbb{\hat O}_{N}$.
It is worth noting that the most straightforward example of an operator with an exact Pauli-based MPO representation (with a bond dimension of one) is a general Pauli string $\hat\Sigma^{\mu}$.
Another useful example which admits an MPO representation with bond dimension one is a projector on a computational basis state
$\ket{s_1,s_2,\dots s_N}$, 
since $\ketbra{s}{s} = (\hat \sigma^{0} + (-1)^{s} \hat \sigma^{3})/2$.\\
 
\paragraph{Clifford \& Magic. ---} \label{s:clifford}
Let us consider the unitary group $\mathcal{U}$ of all unitary matrices acting on $N$ qubits. This group is generated by three elementary gates acting only on a single qubit, namely the Hadamard gate $\hat H$, the phase gate $\hat S$, and the Magic gate $\hat T$; in addition, in order to spread correlations among the system, we need one more generator, the controlled NOT gate $\cnot$ (or CNOT) which acts on two qubits.
Letting these elementary gates act on different qubits, it is possible to implement a generic quantum computation~\cite{BOYKIN2000101,Quantum_Shannon,Drury_2008}.

Among all the unitary matrices, we may consider only those generated by the subset $\{\hat H, \hat S, \cnot\}$; this set, when acting arbitrarily on different qubits, generates the so called Clifford group $\mathcal{U_C}$. This group has the fundamental property to map a Pauli string into another Pauli string (a part from a $\pm 1$ phase) \cite{Gottesman_1997}. Notice that, the CNOT gate, responsible for the entanglement generation, belongs to this group of operators.
In practice, irrespective of the choice of the unitary matrix $\hat C\in\mathcal{U_C}$, although it may generate a significant amount of entanglement when applied to a many-body wave function, it still maintains the inherently low complexity of Pauli string operators.

In fact, what is crucial is the interplay between a generic Clifford unitary and the non-stabilizer generator, aka the Magic gate $\hat T = e^{i\pi(\hat \sigma^{0}-\hat \sigma^{3})/8}$.
To keep the discussion more general, in the following we consider a generic rotation $\hat R^{\mu}(\theta) = e^{-i(\theta/2)\hat\sigma^{\mu}}$ as a local way to inject non-stabilizerness into the many-body wave function (notice that $\hat T \propto \hat R^{3}(\pi/4)$). 
Thanks to the Pauli-based MPO formalism, we can easily account for any Clifford transformation of a Magic gate (or a local rotation).
Let us then assume $\hat C$ an arbitrarily complex Clifford operation on $N$ qubits, and $ \hat R^{\mu}_{l}(\theta)$ a local rotation gate acting on the qubit at position $l$; then we have the following Pauli-based MPO decomposition
\begin{equation}\label{eq:CTC}
\hat C^{\dag} \hat R^{\mu}_{l}(\theta) \hat C 
= \mathbb{\hat T}_{1}\mathbb{\hat T}_{2}\cdots \mathbb{\hat T}_{N}
\end{equation}
where we used the fact that 
$\hat C^{\dag}  \hat \sigma^{\mu}_{l} \hat C = \pm \hat\Sigma^{\boldsymbol{\gamma}}$, with the indices $\boldsymbol{\gamma}=\{\gamma_{1},\dots \gamma_{N}\}$ implicitly depending on both the specific Clifford transformation, and the applied local rotation gate.
The operator-valued MPO has a maximum bond dimension of two and is diagonal in the auxiliary basis. It takes the following form in terms of a controlled Pauli matrix gate (where we omitted the subscript)
\begin{equation}\label{eq:CTC_TN_local}
\includegraphics[width=0.9\linewidth, valign=c]{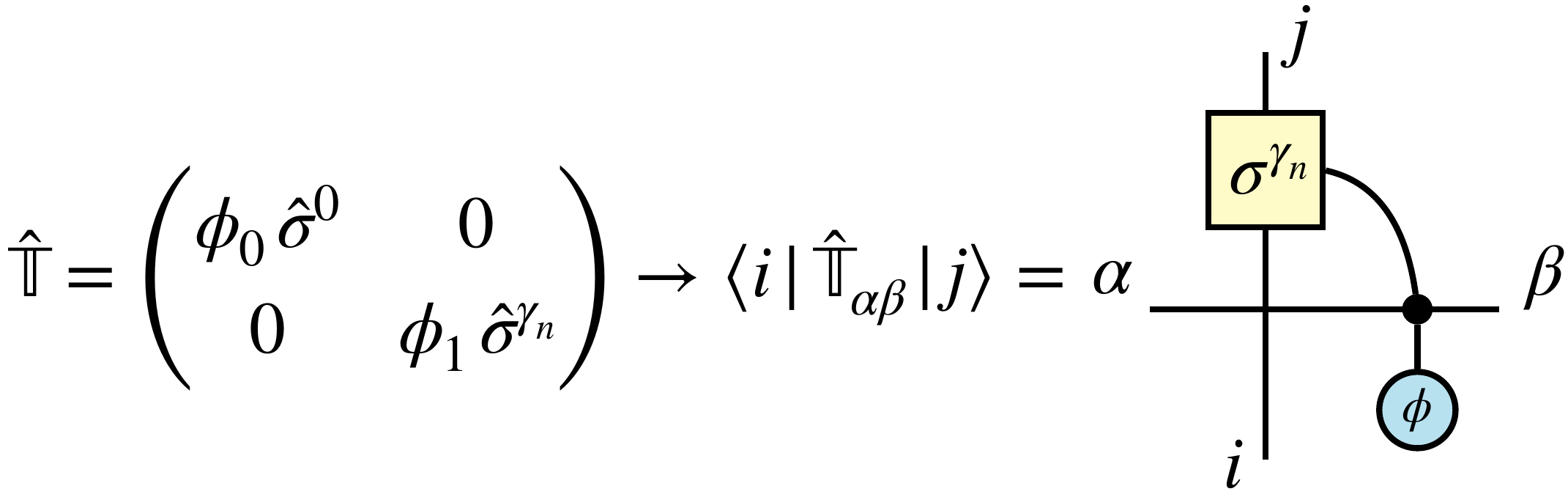}
\end{equation}
with the vector $\phi=\{\phi_0,\phi_1\}=\{c^{1/N},(\mp i\,s)^{1/N}\}$ accounting for the coefficients $c\equiv\cos(\theta/2)$ and 
$s\equiv \sin(\theta/2)$. Notice that in Eq.~\eqref{eq:CTC_TN_local} the indices $\{\alpha,\beta\}$ are spanning the two-dimensional auxiliary space.
The full Tensor Network layer associated to the Clifford-transformed rotation gate reads therefore
\begin{equation}\label{eq:CTC_TN}
\includegraphics[width=\linewidth, valign=c]{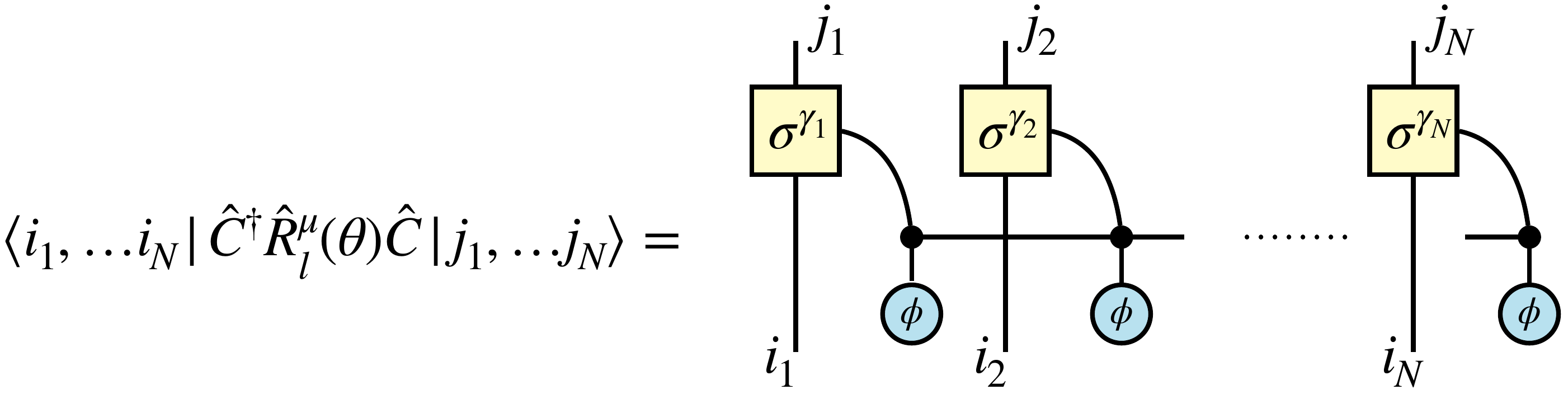}
\end{equation}
From the graphical representation, it becomes immediately evident that whenever $\gamma_n = 0$ for some $n$, the physical and auxiliary spaces become locally disconnected, resulting in a reduction of the complexity of the network. 

Finally, it is worth noting that the auxiliary space linked to each magic layer can be interpreted as an additional qubit, resulting in a two-dimensional TN as represented in Fig.~\ref{fig:stabMPO}. In this case, all local coefficients $\phi$ are collected together to a single auxiliary qubit state $\ket{\theta} = c\ket{0}\mp i s \ket{1}$ accounting for the phase rotation.\\

\paragraph{Objective \& Methods. ---}\label{s:results}
As previously outlined, our objective is to devise an efficient algorithm capable of computing the non-equilibrium dynamics of the expectation value of a local operator, essentially Pauli strings. In other words,
we want to be able to evaluate the following expectation value
\begin{equation}\label{eq:pauli_average}
    \bra{\psi} \hat U^{\dag} \hat \Sigma^{\boldsymbol{\mu}}\hat U \ket{\psi},
\end{equation}
for some short-range correlated state $\ket{\psi}$.
Although it could be a very nontrivial task, any unitary matrix can be decomposed as a series of single local Magic gates separated by arbitrarily deep Clifford circuit \cite{BOYKIN2000101}. 
Therefore, in the following, we will consider our unitary to be decomposed as 
$\hat U = \hat R_{j_{M}} \hat C_{M}\cdots
\hat R_{j_2} \hat C_{2}\hat R_{j_1} \hat C_{1}$,
where $M$ is total number of single-quibit Magic gates entering into the decomposition. Here, we are simplifying the notation 
with the single subscript $j_m$ identifying all parameters of the local rotation $\hat R_{j_m}$.
Now it is clear that, applying this decomposition on the state $\ket{\psi}$ as it is, generally leads to an exponentially complex computation due to the unbounded amount of entanglement which could be injected by each single Clifford layer.

In fact, starting from the rightmost layer, it would be natural to look for the best Clifford ``disentangler'' of $\hat R_{j_1} \hat C_{1} \ket{\psi}$. However, since $\hat R_{j_1}$ differs from the identity matrix only for a single qubit, a very good approximation of the disentangler we are looking for should be given just by $\hat C_{1}^{\dag}$.
Reiterating this procedure through all layers leads to the following \textbf{stabilizer MPO}
representation of the original unitary matrix
\begin{equation}
    \hat U = \mathcal{\hat C} \mathcal{\hat T}_{M}\cdots
    \mathcal{\hat T}_{2}\mathcal{\hat T}_{1},
\end{equation}
with 
\begin{equation}
    \mathcal{\hat T}_{m} =\hat C_{1}^{\dag}\cdots \hat C_{m}^{\dag} \hat R_{j_m} \hat C_{m}\cdots \hat C_{1}
    = \mathbb{\hat T}_{m,1}\mathbb{\hat T}_{m,2}\cdots \mathbb{\hat T}_{m,N},
\end{equation}
for $m\in\{1,2,\dots, M\}$, and the leftmost layer 
$\mathcal{\hat C} = \hat C_{M}\cdots \hat C_{1}$
being just a Clifford unitary. As a consequence, Eq.~(\ref{eq:pauli_average}) reduces to
\begin{equation}\label{eq:sMPO_avg}
    \bra{\psi}  
    \mathcal{\hat T}_{1}^{\dag}\cdots\mathcal{\hat T}_{M}^{\dag}
    \hat \Sigma^{\boldsymbol{\nu}}
    \mathcal{\hat T}_{M}\cdots\mathcal{\hat T}_{1}
    \ket{\psi},
\end{equation}
where $\hat \Sigma^{\boldsymbol{\nu}} = \pm
\mathcal{\hat C}^{\dag}\hat \Sigma^{\boldsymbol{\mu}}\mathcal{\hat C}$.
The significant advantage here lies in the fact that the operation of the stabilizer MPO 
$\mathcal{\hat T}_{M}\cdots\mathcal{\hat T}_{1}$ 
on the state $\ket{\psi}$ typically yields a substantially less entangled state compared to the original unitary process.
In addition, when applied to  the complex conjugate state to compute the expectation value, as depicted in Fig.~\ref{fig:stabMPO}, it is worth noting that, thanks to the network's geometry, it is also possible to contract the entire network horizontally, forming an auxiliary-qubit MPS. This configuration may lead to a much slower growth of the so called ``temporal entanglement'' \cite{PhysRevB.107.L060305,SONNER2021168677,carignano2023temporal} compared to the typical vertical contraction approach. Note that the non-unitary evolution of the auxiliary qubits might lead to a zero state after a finite number of vertical layers. This occurrence is due to the vanishing matrix element associated with the Pauli strings corresponding to a specific configuration of auxiliary qubits.

Let us finally mention that this technique can be exploited in conjunction with the folding scheme where basically Eq.~(\ref{eq:sMPO_avg}) can be evaluated as 
${\rm Tr}(\hat\Sigma^{\boldsymbol{\nu}} \mathcal{\hat T}_{M}\cdots\mathcal{\hat T}_{1}\ketbra{\psi}{\psi}\mathcal{\hat T}_{1}^{\dag}\cdots\mathcal{\hat T}_{M}^{\dag})$.
We can indeed represent the folded TN as
$
\sum_{\boldsymbol{\mu}}
\mathbb{Y}_{1}^{\mu_1}\cdots \mathbb{Y}_{N}^{\mu_N}
\hat \sigma^{\mu_1}\cdots\hat\sigma^{\mu_N}.
$ 
Applying a new layer of the network induces a transformation of each local tensor according to (here we omit the subscripts to simplify the notation)
$
\mathbb{Y}^{\mu} \to 
\sum_{\nu} 
\mathbb{W}^{\mu\nu}\mathbb{Y}^{\nu} 
$
where $\mathbb{W}^{\mu\nu}$, is a diagonal matrix acting on a four dimensional auxiliary space, whose entries are
$\mathbb{W}^{\mu\nu}  = 
\frac{1}{2}{\rm Tr}[\hat\sigma^{\mu}\mathbb{\hat T}\hat\sigma^{\nu}\mathbb{\hat T}^{\dag}] =
{\rm diag}(|\phi_0|^2 \delta_{\mu\nu},\phi_0\phi_1^{*} \Gamma_{\mu\nu}, \phi_0^{*}\phi_1 \Gamma_{\nu\mu},|\phi_1|^2 S_{\mu}\delta_{\mu\nu})
$,
where
$\Gamma_{\mu\nu}=\Gamma^{*}_{\nu\mu}={\rm Tr}(\hat\sigma^{\mu}\hat\sigma^{\nu}\hat\sigma^{\gamma})/2
= i^{\varepsilon_{0\mu\nu\gamma}}\delta_{\mu\oplus\nu\,\gamma}$, 
$S_{\mu}={\rm Tr}(\hat\sigma^{\mu}\hat\sigma^{\gamma}\hat\sigma^{\mu}\hat\sigma^{\gamma})/2 = -(-1)^{\delta_{\mu\gamma}+\delta_{\mu 0}+\delta_{\gamma 0}}$
and $\gamma$ accounts for the Pauli matrix appearing in $\mathbb{T}$ 
(see Eq.~(\ref{eq:CTC_TN_local})). 
In the previous definitions, $\varepsilon_{0\mu\nu\gamma}$ is the Levi-Civita symbol, $\mu\oplus\nu$ is indicating the bit-wise xor between the indices (see Fig.~\ref{fig:stabMPO}(c) for a graphical representation of the folded network).

The strategy outlined in this section can be applied in various scenarios.
\\

\begin{figure}[t!]
    \centering
    \includegraphics[width=0.48\linewidth,valign=c]{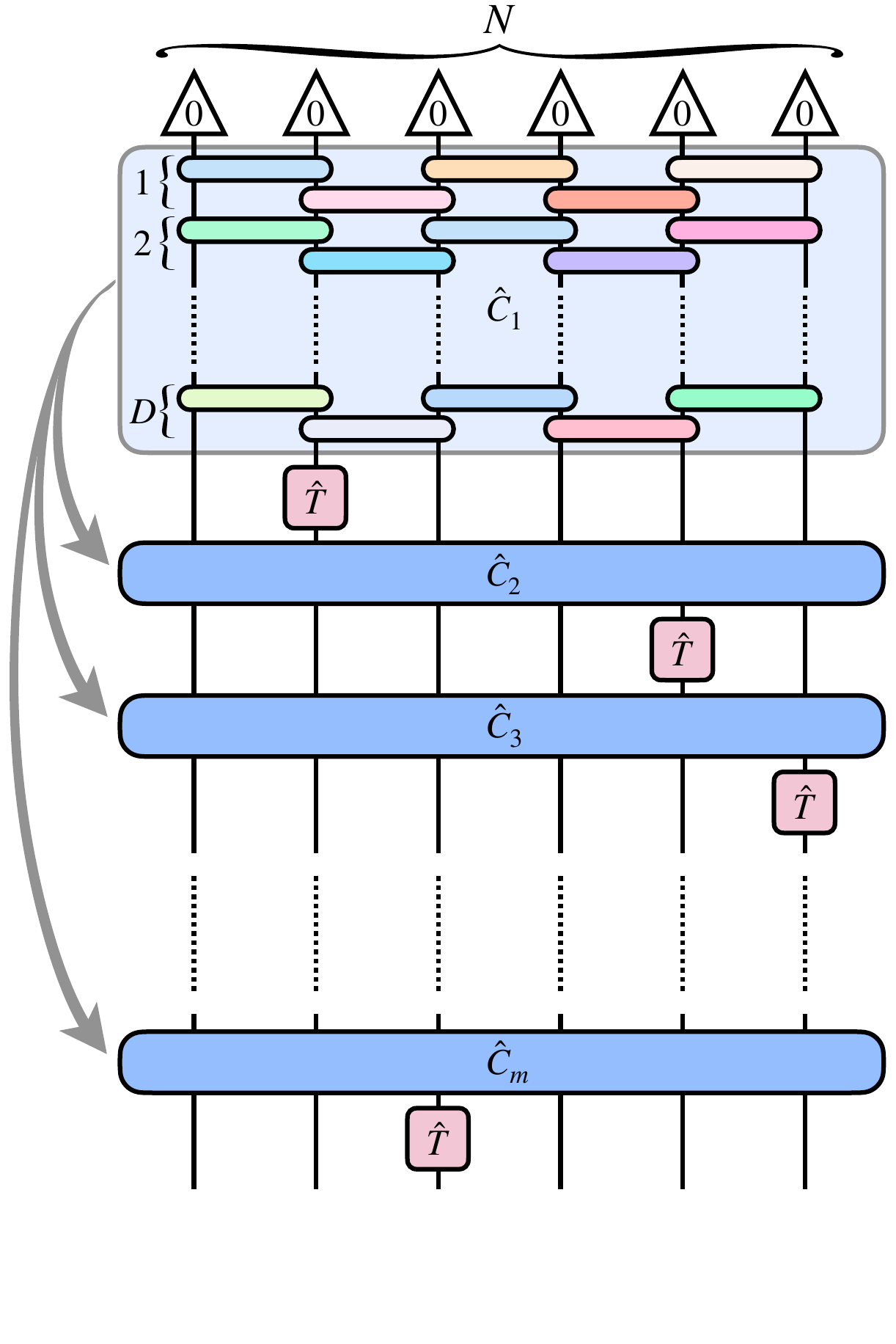}
    \includegraphics[width=0.50\linewidth,valign=c]{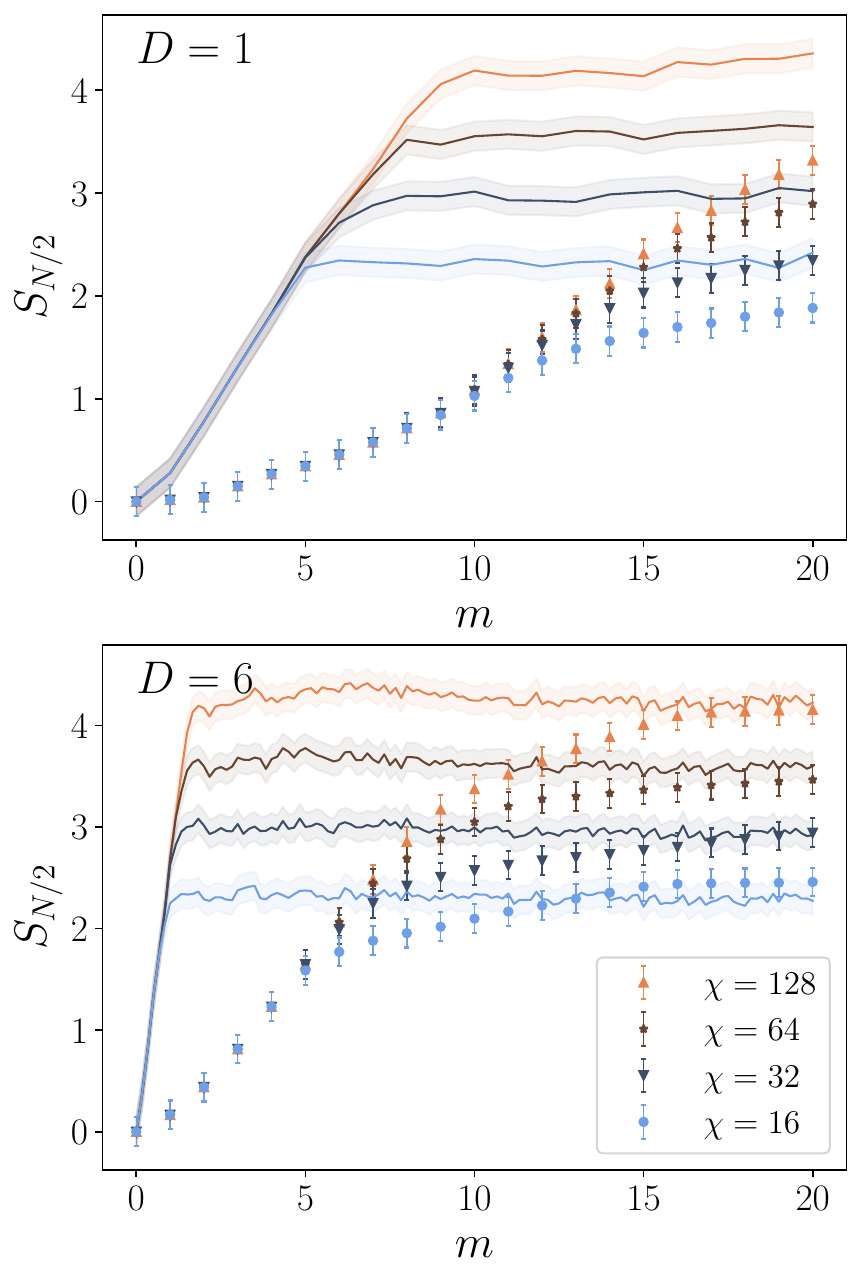}
    \caption{Random Clifford T-doped circuit. \textit{Left panel:} sketch of the evolution. \textit{Right panels:} evolution of the half-chain entanglement entropy of the state evolved according to the stabilizer-MPO formalism (dots) vs the standard full state entanglement entropy (solid lines). The system size is $N=40$.} 
    \label{fig:evo_entanglement}
\end{figure}

\begin{figure}
    \centering
\includegraphics[width=\linewidth]{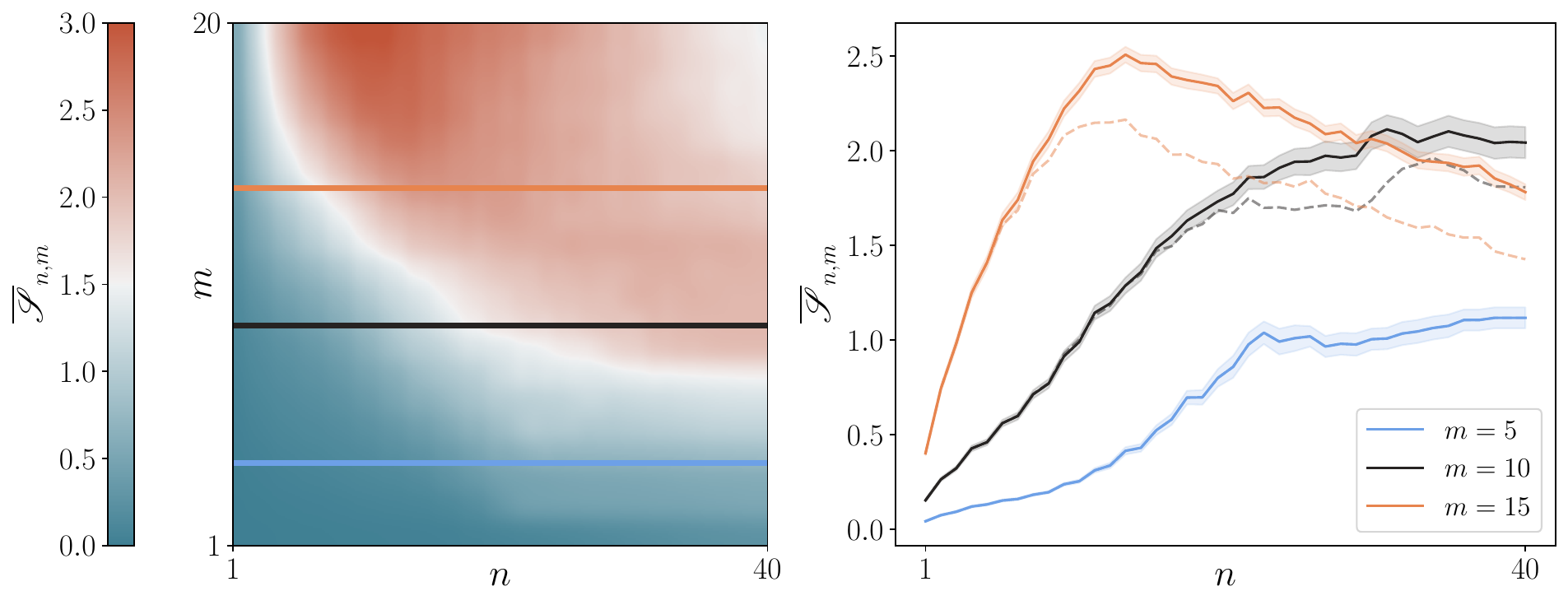}
    \caption{Random Clifford T-doped circuit. \textit{Left panel:} evolution of the average half-chain temporal entanglement entropy $\overline{\mathscr{S}}_{n,m}$, for the observable $\hat{O} = \hat{\sigma}^3_{N/2}$, against the number of sites $n$ and time $m$; parameters $D=1$, bond dimension $\chi=64$. \textit{Right panel:} line plot extracted from left panel for some values of $m$: dashed-line $\chi=32$, solid line $\chi=64$.  }
    \label{fig:temporal entanglement}
\end{figure}

\paragraph{Random Clifford T-doped circuit. ---}
 Our first investigation focuses on random Clifford T-doped circuits. Specifically, we consider circuits 
consisting of $m$ brick-wall shaped random Clifford layers of depth $D$ followed by a $\hat{T}_j$ gate acting on a random qubit $j$ as shown in the left panel of Fig.~\ref{fig:evo_entanglement}.
We apply our protocol to a $N = 40$ qubits system prepared in the state $\ket{0}^{\otimes N}$ subject to $M=20$ intertwined brick-wall Clifford and local T layers. In Fig.~\ref{fig:evo_entanglement} right panels we show the behavior of the entanglement entropy for the half-chain averaged over $50$ realizations, both for $D=1$ and $D=6$ brick-wall layers. Remarkably, by disentangling the state, for a fixed amount of resources $\chi$, the implemented protocol allows to push the time evolution simulation further with respect to the standard sequential application of all the gates (solid lines). Notice that, multiple $\hat T$ gates acting simultaneously on the system can be addressed within the same setup.

Furthermore, as illustrated in Fig.~\ref{fig:temporal entanglement}, we present the evolution of the averaged temporal-entanglement entropy $\overline{\mathscr{S}}_{n,m}$ with respect to the observable $\hat{O}=\hat{\sigma}_{N/2}^3$. For each random realization, $\hat{O}$ has been transformed according to a series of Clifford layers associated with the specific $m$ configuration. To clarify, following the methodology depicted in Fig.~\ref{fig:stabMPO}(b), for each fixed evolution depth $m$, the corresponding left boundary auxiliary vector undergoes systematic horizontal evolution, incrementally adding physical layers from $n=1$ to $n=N$. At each step, the symmetric bi-partited entanglement entropy is computed.
Although in this setup the horizontal contraction does not lead to a significant advantage in terms of the employed resources, it paves the way for further studies in more specific settings.\\

\begin{figure}[t!]
    \centering
    \includegraphics[width=0.5\linewidth]{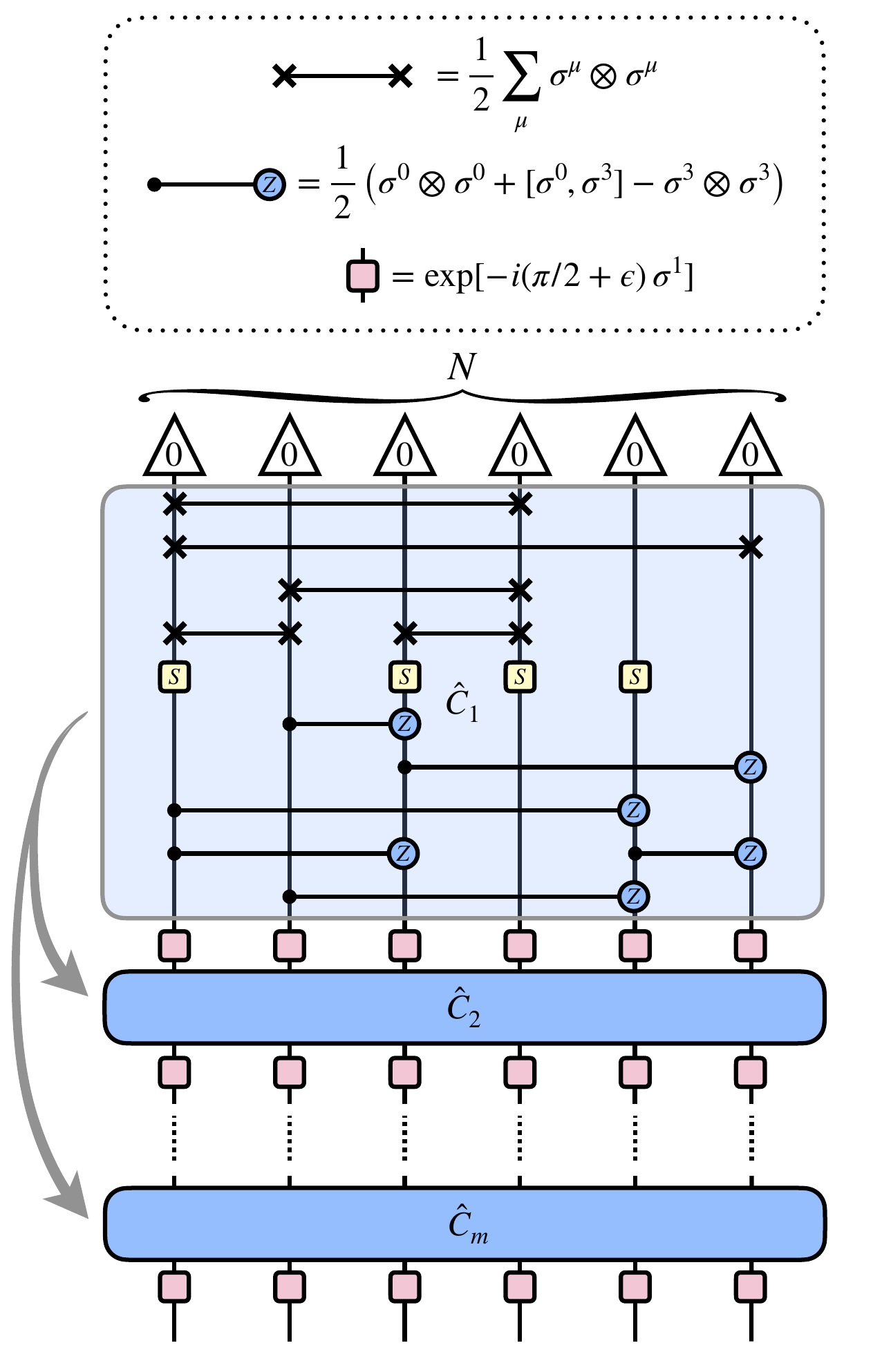}\includegraphics[width=0.49\linewidth]{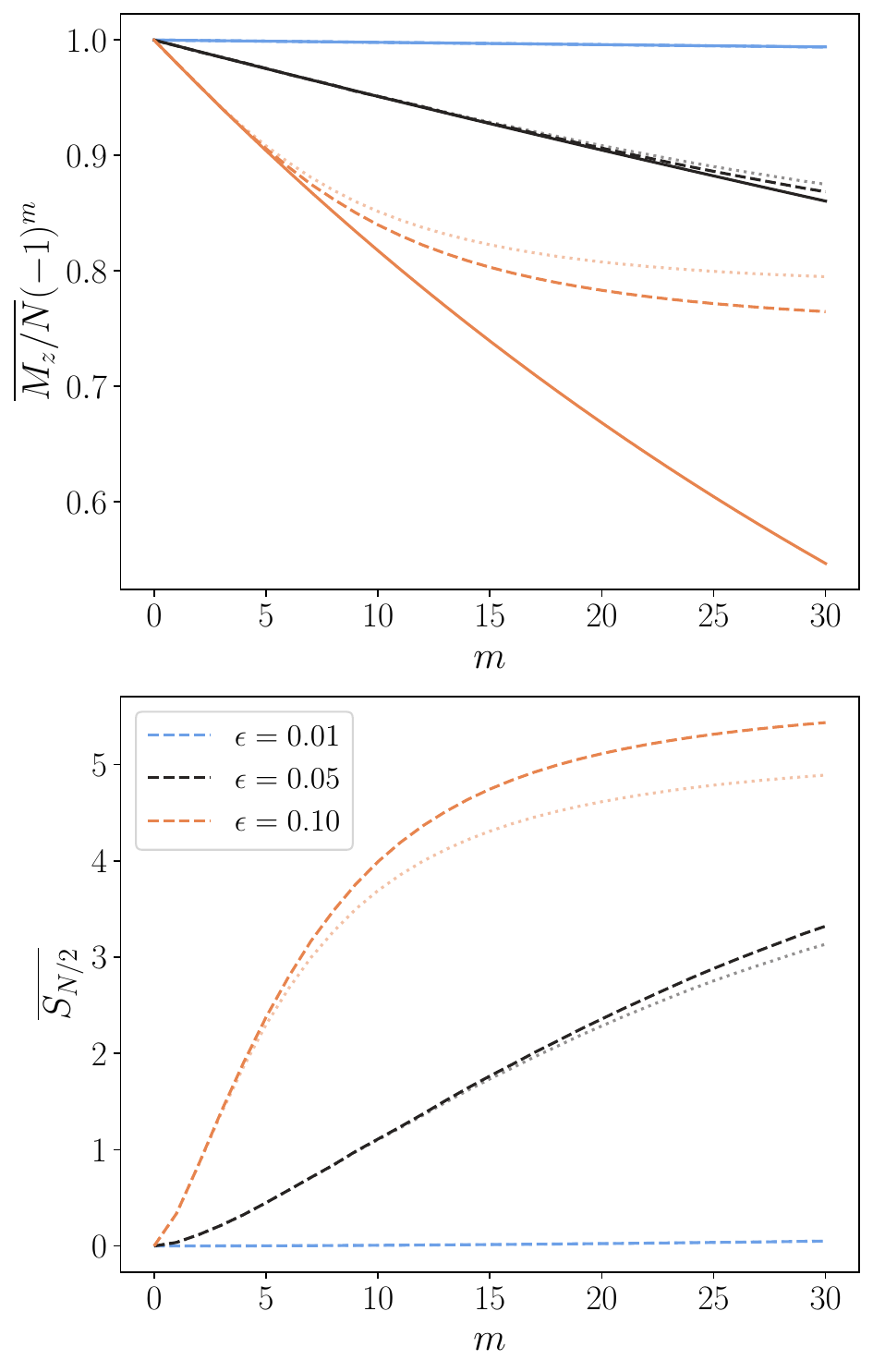}\caption{Random Clifford Floquet Dynamics. \textit{Left panel:} sketch of the evolution. \textit{Top right panel:} evolution of the kicked magnetization for some values of the kick strength $\epsilon$ and $N=40$. \textit{Bottom right panel:} evolution of the half chain entanglement entropy. Solid lines, analytical prediction; dashed lines, bond dimension $\chi=512$; dotted lines, bond dimension $\chi=256$.}
    \label{fig:evo-floquet}
\end{figure}

\paragraph{Random Clifford Floquet Dynamics. ---}
Here we consider the case of a Floquet dynamics induced by the repeated action on the initial state $\ket{0}^{\otimes N}$
of the single period evolution operator
\begin{equation}\label{u1_clifford}
\hat U_{m} = 
\left(\prod_{j=1}^{N}\hat R^{1}_{j}(\pi+2\epsilon) \right)
\hat C_{m}
\end{equation}
where $m$ is indicating the period step, and $\hat C_{m}$
are random $U(1)$-symmetric Clifford gates acting on $N$ qubits; in other words $[\hat C_{m}, \hat M] = 0$ where
$\hat M = \sum_{j}\hat \sigma^{3}_{j} / N$. In fact, this is exactly the scenario where
we basically mimic, in a random Clifford setup, the usual Floquet behavior \cite{PhysRevLett.109.160401,PhysRevX.12.031037,lópez2024exact,PhysRevB.106.134301,PhysRevB.108.L140102,PhysRevResearch.6.013311}. As a matter of fact, the only deviation from the perfect oscillation of the magnetization is induced by a finite value of $\epsilon$. 
In particular, a generic operator of the $U(1)$-symmetric Clifford group can be constructed as~\cite{PRXQuantum.4.040331}
\begin{equation}
    e^{i\phi}
    \left(\prod_{1\leq i<j \leq N} \hat{\rm CZ}^{\nu_{ij}}_{ij}\right)
    \left(\prod_{j=1}^{N} \hat S^{\mu_j}_{j} \right)
    \hat P_n
\end{equation}
with $\mu_j \in \{0,1,2,3\}$, 
$\nu_{ij} \in \{0,1\}$,
, $\phi \in [-\pi,\pi]$, and where $\hat P_n$ is a generic permutation operator acting on $n$ qubits which can be implemented using up to $N(N-1)/2$ swap operators.
In addition $\hat S = {\rm diag}(1,i)$ is the single qubit Phase gate,
and $\hat{\rm CZ} = {\rm diag}(1,1,1,-1)$ is the Controlled-$Z$ gate.
For our purpose the global phase factor is irrelevant so we fixed $\phi=0$.

Let us stress that, due to the highly non-local nature of the Clifford unitary layers $\hat C_m$, this setup is out-of-reach for a standard Time Evolving Block Decimation~(TEBD) algorithm~\cite{Vidal_2004}, while can be easily tackled with our stabilizer MPO approach. 

In Fig.~\ref{fig:evo-floquet} we show the stabilizer MPO evolution of the averaged magnetization and entanglement entropy averaged over different random realisation, and for two different bond dimension. We compare the numerical results with the exact analytical expression of the average magnetization (see~\cite{suppmat}) 
\begin{equation}
    \overline{M_z/N}(m) = (-1)^m{(\cos{2\epsilon})}^m.
\end{equation}
We find a good agreement between the analytical prediction and numerical simulation for small values of $\epsilon$ which could not be verified with standard TN methods.\\

\paragraph{Conclusion \& Outlook. ---} In this work, we have developed a hybrid method that combines the stabilizer formalism with TN techniques to simulate the dynamics of many-body quantum systems. Our stabilizer MPO approach reduces the complexity of unitary evolution by leveraging Clifford transformations to disentangle the quantum state. This allows for efficient computation of expectation values involving Pauli strings and enables longer simulation times compared to standard methods. The stabilizer MPO protocol shows significant potential for optimizing classical simulations of quantum dynamics, offering a promising tool for further research in quantum information science. Further investigations could be focused on finding the optimal disentangler for a MPS, whose existence has been recently pointed out in~\cite{lami2024quantum}, aiming to reduce the volume law of bipartite entanglement entropy to a logarithmic scaling. 
Additionally, it provides a robust framework for benchmarking quantum devices by facilitating accurate comparisons between classical simulations and the performance of quantum hardware.\\

Note Added: A related study, featuring an algorithm where projective measurements of Pauli operators are commuted with Clifford circuits, will be published in the same arXiv posting. The paper is authored by A. Paviglianiti, G. Lami, M. Collura, A. Silva.\\

\paragraph{Acknowledgments. --}
We acknowledge the use of Stim \cite{stim} for the stabilzer formalism operations and ITensor for the TN simulations \cite{itensor}. We are particularly grateful to Guglielmo Lami and Jacopo De Nardis for inspiring discussions, and to Martina Frau, Alessio Paviglianiti, Poetri Tarabunga, Emanuele Tirrito, Marcello Dalmonte for collaborations on topics connected with this work. This work was supported by the PNRR MUR project PE0000023-NQSTI, and by the PRIN 2022 (2022R35ZBF) - PE2 - ``ManyQLowD''.

\bibliography{bib}

\begin{widetext}
\newpage
\section*{SUPPLEMENTAL MATERIAL}

Here, we derive an analytical expression for the decay of magnetization throughout the evolution in the random Clifford Floquet setup. From now on, we denote by $\hat{C}$ a random Clifford operator belonging to the group $\mathcal{U_C} \cap U(1)$, that is a magnetization preserving Clifford unitary. Being $\hat{M}_z =  \sum_{i=1}^N \hat{Z}_j/N$ the magnetization operator, we are essentially interested in evaluating
\begin{align}
    \overline{Z_j}(m) &= 
    \overline{
    \expval{\hat{C}_1^\dag {\hat{R}^{\dag}}
    \dots 
    \hat{C}_m^\dag{\hat{R}^{\dag}}
    \hat{Z}_j\hat{R}\hat{C_m}\dots \hat{R}\hat{C_1}}{0\ ...\ 0}\nonumber}\\ 
    &= \overline{ 
    \Tr\left( \hat{Z_j} \hat{R} \hat{C}_m \dots \hat{R}\hat{C}_1\dyad{0 ... 0}  \hat{C}_{1}^{\dag}\hat{R}^{\dag}
    \dots \hat{C}_{m}^{\dag}\hat{R}^{\dag}\right)}\nonumber\\
    &= \overline{ \mel{Z_j}{ 
    (\hat{R} \otimes \hat{R}^{\ast})
    (\hat{C}_m \otimes \hat{C}^{\ast}_m)
    \dots
    (\hat{R} \otimes \hat{R}^{\ast})
    (\hat{C}_1 \otimes \hat{C}_1^{\ast}) 
    }{0 ... 0\, 0 ... 0} }\label{eq:folded_random_u1}
\end{align}
with $\hat R = \prod_{j=1}^{N}\hat R^{1}_{j}(\pi+2\epsilon) $. Rewriting the discrete parametrization for the generic Clifford as $ 
\hat{C} = \hat{ \rm CZ}{}^{\vec{\nu}} \hat{S}^{\vec{\mu}} \hat{P}$, 
we can graphically represent the quantity we are averaging as shown in Fig.~\ref{fig:folding_randomfloquet}.

\begin{figure}[h!]
    \centering
    \includegraphics[width=\linewidth]{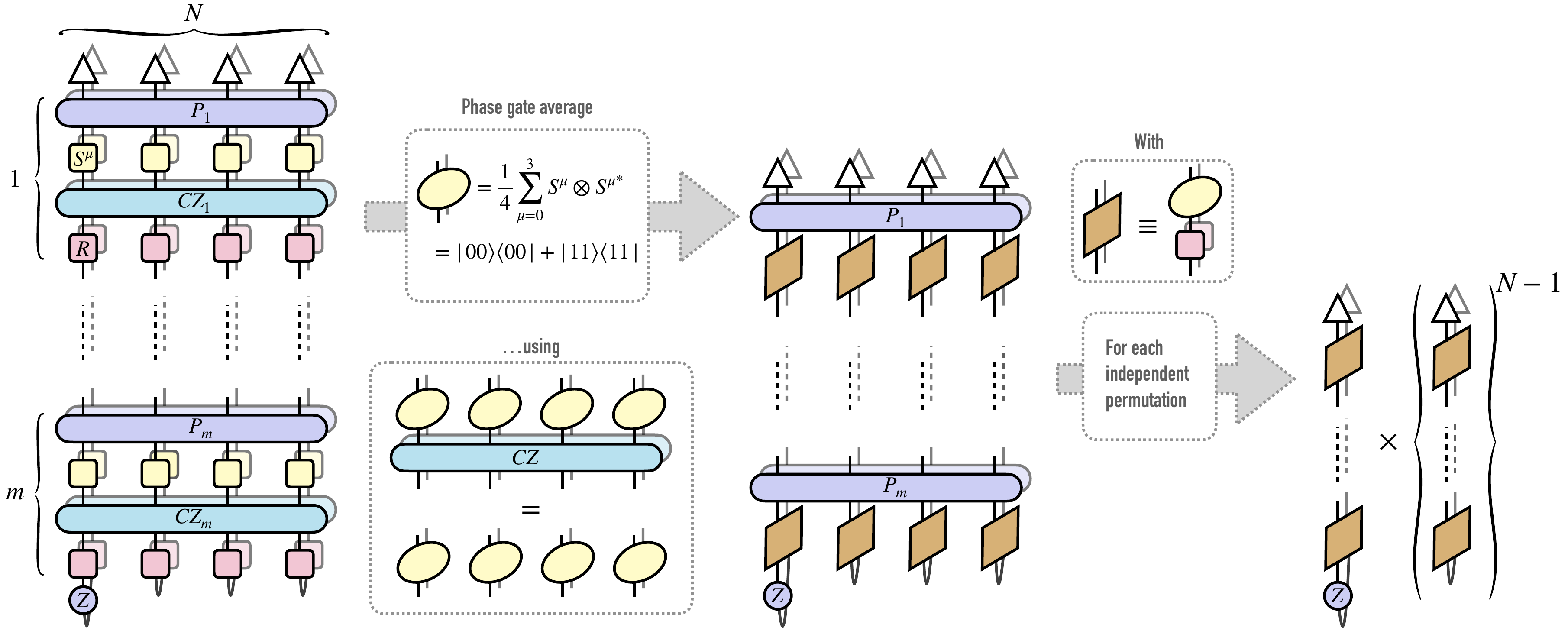}
    \caption{ Sketch of the average over the $U(1)$-symmetric Clifford group, see the text for additional details. 
    \label{fig:folding_randomfloquet}}
\end{figure}

The average in the three terms, i.e. permutations $\hat{P}$, $\hat{S}$ gates and controlled-Z gates, appearing in the decomposition of $\hat{C}$ can be performed independently. Everything gets immediately simpler by noticing that the average of the local $S$ gates reads
\begin{align}
    \hat{\mathcal{S}} = \frac{1}{4} \sum_{\mu=0}^{3} \hat{S}^{\mu} \otimes \hat{S}^{\mu \ast} = \dyad{00} + \dyad{11}.
\end{align}
The correlated action of $\hat{\rm CZ}$ in two replicas of the system is thus trivial no matter the qubits which it is acting on. Indeed,
labeling with $c$ ($c'$) the control qubit of the system (replica), and with 
$t$ ($t'$) the target qubit of the system (replica), we have that
\begin{equation}
\hat{ \rm CZ}_{ct}\otimes \hat{ \rm CZ}_{c't'} \left( \dyad{00} + \dyad{11}\right)_{cc'}\otimes \left( \dyad{00} + \dyad{11}\right)_{tt'} = \left( \dyad{00} + \dyad{11}\right)_{cc'}\otimes \left( \dyad{00} + \dyad{11}\right)_{tt'}.
\end{equation}   
Additionally, each permutation acts identically both on the state and on its copy. Moreover, apart from the application of the $Z_j$ gate, the system is permutationally invariant therefore any permutation brings the same contribution to the average. 

We can therefore define the operator $[ \hat{R}^1(\pi + 2\epsilon) \otimes \hat{R}^{1*}(\pi+2\epsilon)] \hat{\mathcal{S}}$ (brown boxes in Fig.~\ref{fig:folding_randomfloquet}) for which the states $\ket{00} \pm \ket{11}$ are eigenstates with eigenvalues $1$ and $-\cos(2\epsilon)$ respectively. As a consequence, all $N-1$ qubits where $\hat{Z}_j$ does not appear will contribute with $1^m$ for a $m$ layers evolution to the average in Eq.~\eqref{eq:folded_random_u1}, and, the only nontrivial contribution does come from the single qubit on which $\hat{Z}_j$ is acting, such that finally we get
\begin{align}
    \overline{Z_j}(m) = (-1)^m[\cos(2\epsilon)]^m.
\end{align}

\end{widetext}

\end{document}